\documentclass[12pt]{article}
\usepackage{scicite}
\usepackage{soul}
\usepackage{times}
\usepackage{amsmath}
\usepackage{amsfonts}
\usepackage{amssymb}
\usepackage{graphicx}

\usepackage{color}
\usepackage[retain-unity-mantissa=false]{siunitx}
\usepackage{comment}

\newenvironment{sciabstract}{%
\begin{quote} \bf}
{\end{quote}}
\newcommand{\Te}{T_{\textrm{e}}}
\newcommand{\Ti}{T_{\textrm{i}}}
\newcommand{\nugb}{\nu_{\textrm{gb}}}

\title{Electronic Signature of Melting Onset in Polycrystalline Copper at Extreme Conditions}

\author
{Edna R. Toro,$^{1,2\ast}$ 
Tobias Held,${}^{3}$ 
Armin Bergermann,$^{2}$ 
Megan Ikeya,$^{2}$\\
Maximilian Maigler,${}^{4}$ 
Eric R. Sung,$^{2}$ 
Jochen Schein,${}^{4}$ 
Dirk O. Gericke,$^{5}$\\
Mianzhen Mo,$^{2}$
Baerbel Rethfeld,$^{3}$
Siegfried H. Glenzer,$^{2}$\\
Benjamin K. Ofori-Okai$^{2\ast}$\\
\\
\normalsize{$^{1}$Department of Mechanical Engineering, Stanford University, Stanford, CA, USA}\\
\normalsize{$^{2}$SLAC National Accelerator Laboratory, Menlo Park, California, 94025, USA}\\
\normalsize{$^{3}$Department of Physics and State Research Center OPTIMAS,}\\
\normalsize{RPTU University Kaiserslautern-Landau, Kaiserslautern, Germany,}\\
\normalsize{$^{4}$Bundeswehr University Munich,}\\
\normalsize{Werner-Heisenberg-Weg 39, 85579 Neubiberg, Germany,}\\
\normalsize{$^{5}$Centre for Fusion, Space and Astrophysics,}\\
\normalsize{Department of Physics, University of Warwick,}\\
\normalsize{Coventry CV4 7AL, United Kingdom}\\
\normalsize{$^\ast$Correspondent author; E-mail: etorogar@stanford.edu,}\\
\normalsize{benofori@slac.stanford.edu}
}

\begin{document}
\maketitle
\baselineskip24pt

\date{}

\begin{sciabstract}
Ultrafast melting is fundamentally a structural transition of the ionic lattice, but this rearrangement also reshapes the electronic properties by changing the energy landscape and scattering mechanisms.
Although the electrons react almost instantaneously, it is not {\it a priori} clear how much lattice disorder is required for a significant optical response.
Here, we show that the onset of melting already produces a clear electronic signature in polycrystalline copper. 
Using single-shot terahertz time-domain spectroscopy on thin films excited over a wide range of laser fluences, we infer the transient conductivity during the first picoseconds after excitation.
The data, supported by two-temperature molecular-dynamics simulations, show that electron transport is substantially limited by grain-boundary scattering before melting and that melting strongly suppresses this process.
As melting begins at these interfaces, we observe a transient increase in the conductivity that directly marks the onset of the phase transition.
More broadly, these results show that the ionic and electronic relaxation stages are closely coupled in nonequilibrium laser-driven matter and that optical measurements can resolve distinct stages of melting.

\end{sciabstract}


\section*{Introduction}
\label{sec:intro}

Ultrafast melting of laser-excited metals is a fundamental phase transition in which the loss of long-range order drives major changes in thermodynamic and transport properties.
Since the evolving structure changes the environment of the conduction electrons, observables correlated with electronic 
behavior, such as reflectivity, can distinguish between solid and fully molten states~\cite{Wellershoff1999, Kandyla2007}.
However, the stage of melting at which these changes occur has not yet been identified.
This question is of particular interest because experiments can resolve the melting process even for ultrafast excitations where the lattice is heated by energy transfer from the hot electrons \cite{maigler2024atomistic, MoCu}.
Observables that can identify both the onset and completion of melting are therefore particularly useful.

Structural probes such as x-ray and electron diffraction have shown that crystalline order can persist for several picoseconds after femtosecond excitation before melting begins~\cite{levy2009x, mahieu2018probing, jourdain2018electron, jourdain2021ultrafast, mo2021ultrafast, dorchies2023nonequilibrium, maigler2024atomistic, White2025, MoCu}.
This persistence of order suggests that, during the earliest stages of disordering, the electronic properties may also retain memory of structural features inherited from the initial solid.
Consistent with this picture, density-functional-theory and molecular-dynamics calculations indicate that conductivity in strongly driven metals depends not only on electron and ion temperatures, but also on the instantaneous atomic arrangement and electronic structure~\cite{witte2018observations, schorner2022ab}.

\begin{figure*}[b]
  \centering
  \includegraphics[width=0.9\textwidth]{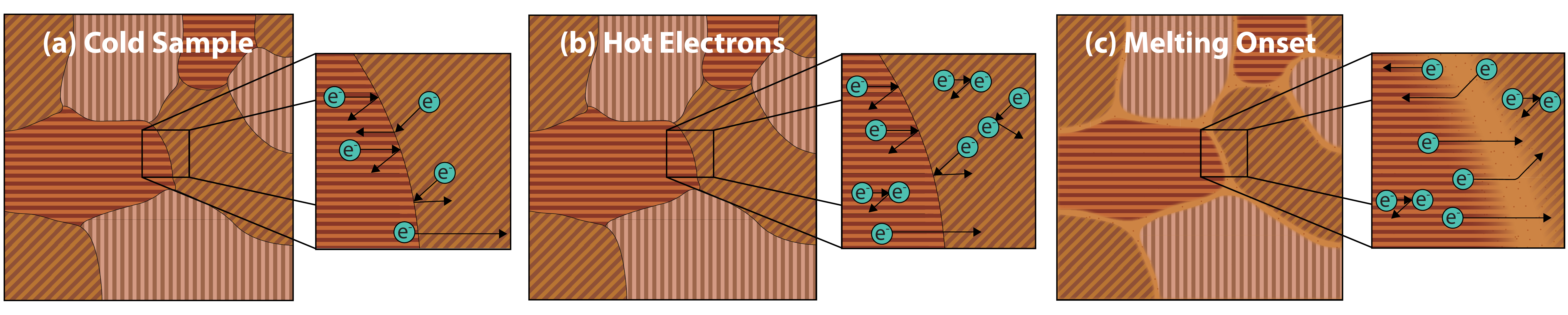}
  \caption{
  {\bf Schematic of the proposed physical picture in laser-excited polycrystalline copper.}
  (a) In the cold film, intact grain boundaries act as transport-limiting interfaces and suppress conductivity.
  (b) Immediately after excitation, the hotter electrons scatter more frequently from bound electrons, while the grain-boundary contribution remains active since the polycrystalline structure is still preserved.
  (c) As melting begins preferentially at grain boundaries, a liquid layer forms between the grains, thereby removing the additional scattering mechanism.
  }
  \label{fig:sketch_gb}
\end{figure*}

Polycrystalline thin films provide a particularly promising system for testing this idea.
Under ambient conditions, grain boundaries are known to reduce the conductivity of metallic films when the grain size becomes comparable to the electronic mean free path \cite{Mayadas1970, chopra1970thin, palasantzas1998surface, mercier2002introduction, wu2004influence, sun2010surface, chawla2011electron}.
Recent structural studies further suggest that, in strongly driven polycrystalline metals, melting is initiated preferentially at grain boundaries and other disordered regions before propagating into the grains \cite{Assefa2020, Antonowicz2024}.
Together, these observations motivate the physical picture illustrated in Fig.~\ref{fig:sketch_gb}: if grain boundaries both limit long-range transport and melt first, their disappearance may provide an electronic signature of the onset of melting.

Despite this connection, studies of the ultrafast response in warm dense matter have not considered how nanometer-scale structure inherited from the cold target material contributes to electron damping, focusing instead on electron-electron and electron-ion scattering~\cite{chen2021, ofori2024al}.
Probes based on optical or X-ray pulses may not capture this effect, as they are not sensitive to the collective motion of free electrons.
How long the initial grain structure influences transport properties, and whether its disappearance can be resolved experimentally, remains largely unexplored.

Here, we investigate the link between structural changes and electronic properties in thin polycrystalline copper films excited by femtosecond laser pulses.
Using single-shot terahertz time-domain spectroscopy (THz-TDS), we infer the transient DC conductivity over a wide range of energy densities and pump-probe delays during the first picoseconds after excitation.
The long wavelength and oscillation period of THz radiation makes it an ideal probe as it is sensitive to disorder and transport over nanometer length scales~\cite{Walther2007}, unlike optical or X-ray probing which are sensitive to transport properties over much shorter length scales.
We interpret the measurements using a Drude-based conductivity model supported by two-temperature molecular-dynamics simulations.
We find that, before melting, the conductivity is substantially reduced by scattering at grain boundaries inherited from the initial film for all probed electron temperatures.
As melting begins, this transport-limiting process disappears rapidly, producing a transient conductivity increase that marks the onset of the phase transition.
These results identify the transient conductivity increase as an electronic signature of melting onset in polycrystalline metals, with corresponding consequences for the optical response.

\section*{Experiment}
\label{sec:exp}

\begin{figure*}[tb]
\begin{center}
  \includegraphics{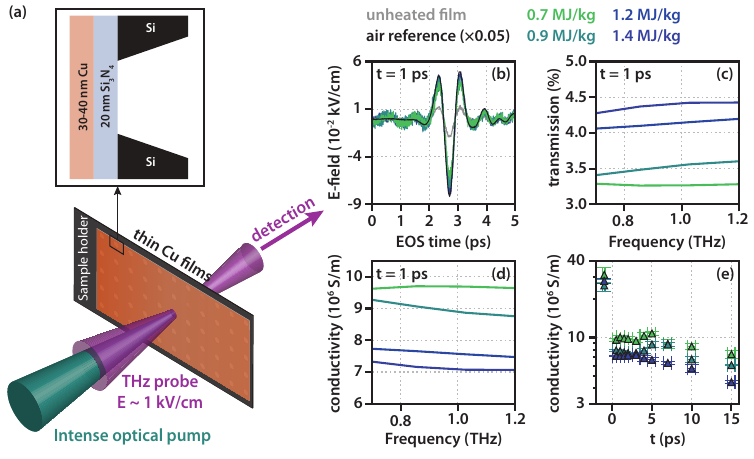}
\end{center}
  \caption{{\bf Experimental details and representative data.} 
  (a) Schematic of the target geometry showing the thin copper film, optical pump and THz probe.
  (b) Representative waveforms of THz pulses transmitted through 30\,nm films probed 1\,ps after laser excitation at different energy densities, labeled with different shades of green (0.7 and 0.9 MJ/kg) and blue (1.2 and 1.4 MJ/kg).
  The gray waveform corresponds to the unheated film and the black waveform is the air reference, scaled by 0.05, taken after ablation. 
  (c) Transmission spectra obtained by Fourier transform of the waveforms in panel (b), normalized by the spectrum of the air reference.
  (d) Frequency-domain conductivity inferred from transmission measurements using Eq.~\eqref{eq:tinkham}. 
  (e) Time-resolved electrical conductivity obtained by averaging conductivity spectra over 0.8 to 1.2 THz for different energy densities and pump-probe delays.
  }
  \label{fig:experiment}
\end{figure*}

Our measurements were performed with a custom tabletop single-shot THz-TDS system, adapted from previously reported designs~\cite{ofori2018toward, ofori2024dc, ofori2024al}.
Figure~\ref{fig:experiment}(a) shows a schematic of the interaction plane.
Polycrystalline copper films with thicknesses of $30.5\pm 3.0$~nm and $39.3\pm 2.7$~nm were heated by 50 fs FWHM, $\lambda=400$~nm laser pulses.
The highest probe-averaged fluence was 158.1$\pm$0.5~mJ/cm$^2$, corresponding, under the assumption of constant reflectivity, to an increase in energy density of $\rho_{\varepsilon} = 1.81\pm0.22$~MJ/kg.
The sample was probed using THz pulses generated by optical rectification of $\lambda=800$~nm pulses in a N-benzyl-2-methyl-4-nitroaniline (BNA) crystal~\cite{zhao2019efficient, tangen2021comprehensive}.
The transmitted THz pulses were recorded by single-shot electro-optic sampling using a reflective echelon~\cite{Minami2013, ofori2018toward, chen2021, ofori2024dc, ofori2024al}. 

The thin copper films were produced by Silson Ltd.~(UK) by sputtering onto a 20\,nm silicon nitride (Si$_\textrm{3}$N$_4$) layer.
Both layers were supported by a 360\,µm thick silicon wafer.
As shown previously, the Si$_3$N$_4$ layer has a negligible influence on the extracted conductivity owing to its minimal THz absorption~\cite{ofori2024al}.
After deposition, the silicon wafer was back-etched to produce a grid of 600\,µm $\times$ 600\,µm square windows with a 3\,mm pitch.
As the transverse size of the THz pulse is $\sim$ 1 mm diameter, the un-etched silicon acted as a mask, ensuring that the detected THz signal corresponds only to laser-heated material. 

Figure~\ref{fig:experiment}(b) shows representative time-domain waveforms of THz pulses transmitted through an unheated film and through films driven to different energy densities and probed $1$\,ps after excitation.
A hole without a sample was used to measure a reference waveform and determine the absolute transmission.
The transmitted THz field amplitude increases with energy density, suggesting reduced conductivity relative to the cold sample.

We quantified the frequency-dependent transmission coefficient $\tilde{t}\left(\omega\right)$ by Fourier transforming the measured waveforms and normalizing to the reference spectrum.
Figure~\ref{fig:experiment}(c) displays the magnitude of the transmission coefficients corresponding to the waveforms shown in Fig.~\ref{fig:experiment}(b).
The conductivity spectra $\tilde{\sigma}\left(\omega\right)$ were calculated using the Tinkham formula~\cite{tinkham}
\begin{equation}
\tilde{\sigma}\left(\omega\right) = \frac{\tilde{n}_1+\tilde{n}_2}{Z_0 d} \left(\frac{1}{\tilde{t}(\omega)} - 1\right)\,,
\label{eq:tinkham}
\end{equation}
where $\tilde{n}_1 (\omega) = 1$ and $\tilde{n}_2 (\omega)=2.7$ are the refractive indices of the media surrounding the film, air and Si$_\textrm{3}$N$_4$, respectively. $Z_0=377~\Omega$ is the impedance of free space, and $d$ is the thickness of the film.
Figure \ref{fig:experiment}(d) shows the conductivity spectra resulting from the transmission in Fig.~\ref{fig:experiment}(c).

As the extracted conductivity varies only weakly across the spectrum, we use its average value from 0.8 to 1.2~THz, where our THz spectral amplitude is largest, as an estimate of the DC limit.
The resulting temporal evolution of the conductivity is shown in Fig.~\ref{fig:experiment}(e) for energy densities representative for our study.
These time traces form the basis for our analysis.

\section*{Modeling}
\label{sec:modeling}

In noble metals, electrical transport is dominated by electrons in the $sp$ band, while $d$ electrons do not contribute significantly to conduction~\cite{Ndione2022}.
Thus, within a Drude-model framework, the DC conductivity is given by
\begin{equation}
\sigma_0 = \frac{n_{sp} e^2}{m_{e} \nu_{\textrm{tot}}},
\label{eq:Drude_simp}
\end{equation}
where $e$ and $m_{e}$ denote the electron charge and mass and $n_{sp}$ is the density of $sp$ electrons.
We decompose the total damping rate, $\nu_{\textrm{tot}}$, into electron-electron ($\nu_{\textrm{ee}}$), electron-ion ($\nu_{\textrm{ei}}$), and grain-boundary ($\nugb$) scattering contributions
\begin{equation}
  \nu_{\mathrm{tot}}(\Te, \Ti, t)  = \nu_{\textrm{ee}} + \nu_{\textrm{ei}} + \nugb,
  \label{eq:totalfrequencymathiesen}
\end{equation}
following Matthiessen’s rule, where $\Te$ and $\Ti$ are electron and ion temperatures, respectively.

For electron-electron scattering, only collisions between $4sp$ and $3d$ electrons limit the conductivity.
Following Refs.~\cite{Fourment2014, Ndione2022}, we model 
the electron-electron scattering rate as
\begin{equation}
  \nu_{\textrm{ee}}(\Te, t)= A \, n_d(\Te, t) \, n_{d}^{\mathrm{holes}}(\Te, t),
\label{eq:sp_d_scattering}
\end{equation}
where $n_d(\Te, t)$ is the density of $d$ electrons and $n_{d}^{\mathrm{holes}}(\Te, t) = n_d(0\,\mathrm{K}, t) - n_d(\Te, t)$ is the density of $d$-band holes.
The coefficient $A$ for copper will be determined by our measurements.
For copper under ambient conditions, the $3d$ band is nearly fully occupied; hence this contribution is only relevant at elevated electron temperatures.

With increasing electron temperature, carriers are redistributed from the $3d$ into the $4sp$ band, affecting the conductivity through two competing mechanisms.
A higher $sp$ carrier density tends to increase conductivity directly in Eq.~\eqref{eq:Drude_simp}, but the accompanying creation of $d$-band holes enhances the electron damping through scattering as captured by Eq.~\eqref{eq:sp_d_scattering}.
The corresponding carrier densities are obtained from the partial electronic density of states and Fermi occupation.
These carrier densities are also explicitly time dependent due to the evolving sample volume.


We approximate the electron-ion scattering rate as $\nu_{\mathrm{ei}}=B\Ti$~\cite{ashcroft1976solid}, with $B$ fixed by the room-temperature conductivity of bulk copper $\sigma(\SI{300}{\kelvin})= 5.8\times10^7$\,S/m~\cite{matula1979electrical}.
The grain-boundary scattering (GBS) rate depends directly on the nanostructure of the film.
The resulting total conductivity therefore depends on electron temperature through band occupation, on ion temperature through electron-ion scattering, and on sample structure through the carrier densities and GBS.

We obtain the temporal evolution of the electron and ion temperature by solving the two-temperature model (TTM)~\cite{Anisimov1974, Rethfeld2017} (see Materials \& Methods for details).
In parallel, TTM coupled with molecular dynamics (TTM-MD) simulations provide the evolving atomic structure, including the melting onset, loss of crystalline order, and film expansion~\cite{Schaefer2002,Ivanov2003}.
The TTM-MD results are used in two ways.
First, they provide the evolving film thickness and density used in the conductivity analysis.
Second, we use the simulated fraction of atoms in a face-centered cubic (fcc) environment as an indicator of the progression from the initial lattice to a largely molten state.
This interpretation is supported by previous benchmarking against ultrafast electron diffraction measurements~\cite{MoCu,maigler2024atomistic}.

\section*{Results and Discussion}
\label{sec:results}

\subsection*{Parameter extraction and scattering hierarchy}

We first use the cold-film conductivity and the full data set to constrain the free parameters of the transport model.
From the unpumped sample, we obtain $\sigma_0=(2.9\pm0.2)\times10^7$~S/m from the THz measurement, in agreement with the independent four-point-probe value of $(2.9\pm2.0)\times10^7$~S/m.
Ascribing this reduced conductivity relative to bulk copper to GBS, we estimate $\nu_{\mathrm{tot}}(300\,\mathrm{K}) = \nu_{\mathrm{ei}}(300\,\mathrm{K}) + \nugb$, which yields a cold-film GBS rate of $\nugb = (4.2 \pm 0.6)\times10^{13}\,\mathrm{s^{-1}}$, averaged over the films used in this study. The coefficient $A$ in Eq.~\eqref{eq:sp_d_scattering} is then obtained from a global fit to the measured conductivity over all delays and energy densities, yielding $A=(2.2\pm0.2)\times10^{-44}\,\mathrm{m^6\,s^{-1}}$.
With these parameters fixed, the model determines the relative importance of grain-boundary, electron-ion, and electron-electron scattering across the experimentally accessed state space.

\begin{figure*}[t]
  \centering
  \includegraphics[width=0.9\textwidth]{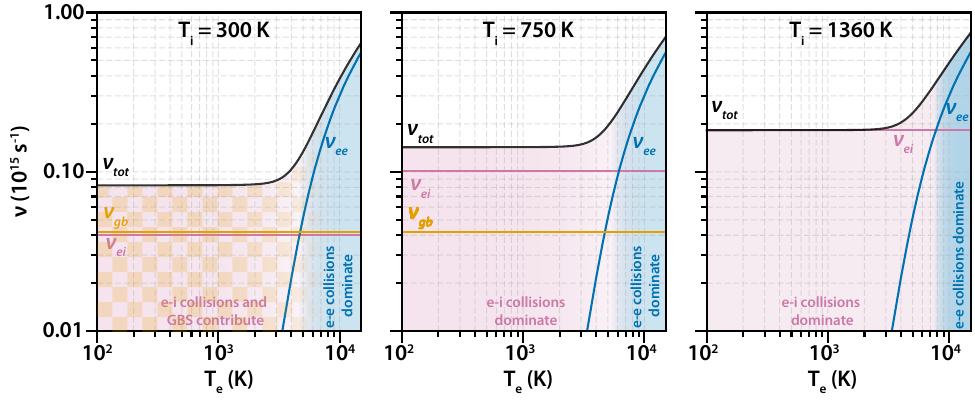}
  \caption{
  {\bf Scattering hierarchy in copper established by the measurements}. For different ion temperatures above and below the melting temperature, $T_{M}=1358$ K,
  the total damping rate (black lines) and its decomposition into electron-ion (pink line), electron-electron (blue line), and grain-boundary contributions (orange line) are shown as functions of electron temperature for three ion temperatures between room temperature and the melting point. The colored regions in the background indicate the dominant damping mechanism (checkered for equal contributions).
  }
  \label{fig:frequencies_modeling}
\end{figure*}

The resulting scattering hierarchy is summarized in  Fig.~\ref{fig:frequencies_modeling}.
For a lattice at room temperature, grain-boundary and electron-ion scattering make comparable contributions to the damping rate, as illustrated by the checkered region.
By contrast, electron-electron scattering is negligible in the cold film but rises rapidly with electron temperature.
It exceeds the grain-boundary contribution above $\Te\sim$ \SI{5000}{\kelvin} and becomes the dominant channel at $\Te\sim$ \SI{10000}{\kelvin}, indicated by the blue region in all figures.
For solid samples with elevated $\Ti$, electron-ion collisions dominate the damping, though GBS still has a noticeable impact, represented by the pink regions.
Of course, GBS no longer contributes in a molten sample with $\Ti > T_{M}=1358$ K.
This hierarchy suggests that the clearest transport signature of grain-boundary removal should appear at intermediate excitation, where $\nugb$ remains significant but melting still occurs within the measurement window.

\subsection*{Conductivity evolution during the melting process}

\begin{figure*}[tb]
  \centering
  \includegraphics[width=0.99\textwidth]{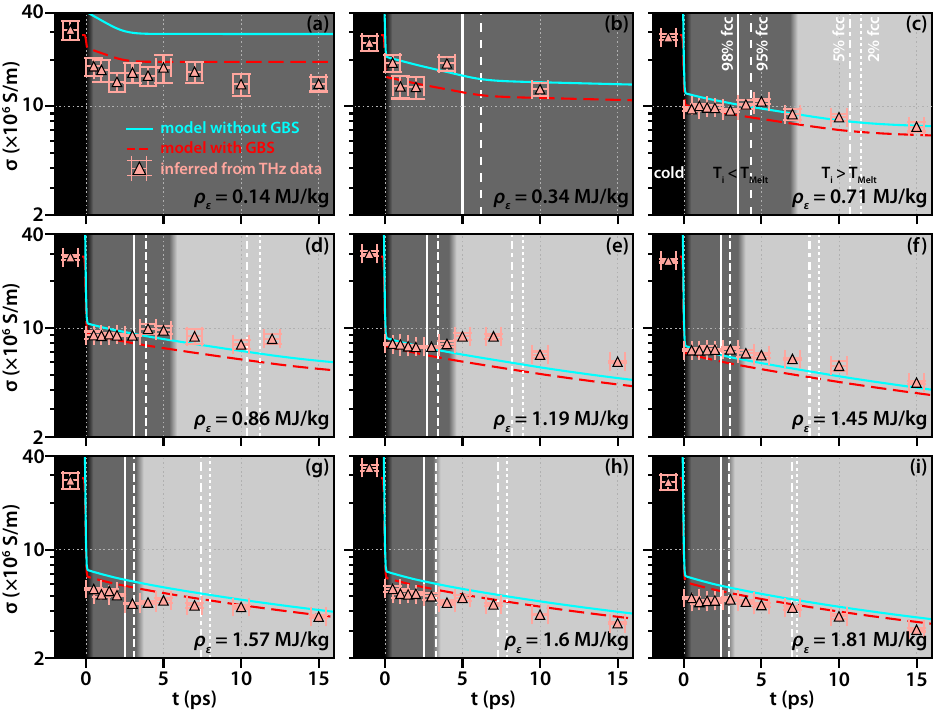} 
  \caption{{\bf Time-dependent conductivity of copper films driven to different energy densities.}
  (a)-(i) Symbols show data inferred from THz-TDS and curves show simulated results.
  Red dashed curves include grain-boundary scattering, while the cyan solid curves do not, i.e. $\nugb=0$.
  Background shading indicates whether the ion temperature predicted by TTM is below or above the melting temperature.
  Vertical lines mark selected values of the remaining fraction of atoms in an fcc environment and serve as structural markers for the progression of melting.
  }
  \label{fig:conductivity_over_time}
\end{figure*}

Figure~\ref{fig:conductivity_over_time} compares the DC conductivity inferred from the THz-TDS measurements with Drude-model calculations for different energy densities.
For each excitation, we perform two calculations: one including all three scattering channels in Eq.~\eqref{eq:totalfrequencymathiesen} and one excluding GBS.
The background shading and vertical markers indicate the thermal and structural evolution predicted by the TTM-MD simulations.
Specifically, the vertical lines mark selected values of the simulated fraction of atoms in an fcc environment, which serve as reference points for the onset and completion of melting.

For all energy densities, the conductivity drops rapidly on the timescale of the pump pulse.
The model reproduces this initial decrease, which reflects the sharp increase in electron scattering in the hot-electron state immediately after excitation.
At the lowest energy density, for which the simulations indicate that the film remains solid-like throughout the measurement window, the initial decrease in conductivity is followed by an extended plateau.
This behavior is typical for laser-heated noble metals~\cite{chen2021} and reflects a transient balance between the decreasing electron-electron contribution as the electrons cool and the increasing electron-ion contribution as the ions heat~\cite{Ndione2022,ndione2024stable}.
In this regime, the inferred conductivity is described well by the calculation that includes GBS, indicating that the transport remains sensitive to the polycrystalline nanostructure inherited from the initial film.

For $\rho_{\varepsilon}>0.14$ MJ/kg, where the simulations indicate that melting begins within the measurement window, a different behavior emerges.
After the initial drop, the conductivity exhibits a modest recovery on picosecond timescales before decreasing again at later times.
At early delays, the data remain close to the calculation that includes GBS, whereas later they shift toward the calculation without GBS.
This crossover occurs at delays that coincide with the simulated onset of melting, where the fraction of atoms in an fcc lattice drops below 95\%, within the experimental and modeling uncertainties  (See supporting information). This is most clearly observed for $\rho_{\varepsilon}<1.5$ MJ/kg, where $\Ti$ remains below $T_{M}$ for several picoseconds after the simulated onset of melting, and the conductivity increases.
We therefore interpret the recovery as an electronic signature of melting onset: as heterogeneous melting begins at grain boundaries and other disordered interfaces~\cite{Assefa2020,Antonowicz2024}, the associated transport-limiting structural contribution disappears.
The subsequent gradual decay in conductivity can be attributed to film expansion after melting begins~\cite{chen2018interatomic}.

For $\rho_{\varepsilon}>1.5$ MJ/kg, the conductivity decreases nearly monotonically after the initial drop.
Although the simulations predict earlier melting in this regime, the transport signature of grain-boundary removal is less pronounced than for lower energy densities.
This trend is consistent with the scattering hierarchy shown in Fig.~\ref{fig:frequencies_modeling}:
at intermediate energy densities, the disappearance of GBS during melting produces an observable change in the total damping rate, whereas at the highest energy densities the same effect is increasingly masked by stronger electron-electron scattering.
In addition, the nucleation rate increases rapidly with lattice superheating, so at higher energy densities melting is not expected to initiate exclusively at grain boundaries~\cite{Antonowicz2024,Rethfeld2002melt}.

\section*{Conclusion}
\label{sec:Conclusion}

We used single-shot THz time-domain spectroscopy to track the transient conductivity of thin polycrystalline copper films.
We excited these targets with femtosecond laser pulses over a broad range of energy densities and probed the conductivity evolution on the few to tens of picosecond timescales.
Analysis based on a Drude model supported by TTM-MD simulations shows that the conductivity cannot be understood from electron and ion temperatures alone.
Instead, at early times, the transport properties still reflect the film's initial polycrystalline structure that manifests as a substantial grain-boundary scattering contribution.

When melting is induced, the conductivity exhibits a transient recovery after the initial drop due to electron heating.
The timing of this recovery coincides with the simulated onset of structural disordering and is consistent with the rapid disappearance of transport-limiting grain boundaries.
Thus, the data show that electrical transport in strongly driven matter tracks the evolving ionic structure almost instantaneously.
In this sense, the conductivity provides an electronic and optical signature of the melting onset.

More broadly, our results show that electrical properties in strongly driven matter are highly sensitive to structural dynamics.
For ultrafast laser-heated metals, transport models based on thermodynamic state variables alone may therefore miss an important structural contribution until the material becomes effectively homogeneous.
An analogous sensitivity to transient structural heterogeneity may also arise in dynamically compressed matter, where defect generation, fragmentation, or partial melting can transiently create transport-limiting interfaces.
Accounting for this structural degree of freedom should improve both the interpretation of ultrafast conductivity measurements and the modeling of transport during the formation of warm dense matter.

\newpage
\section*{Materials and Methods}

\subsection*{Setup  for the THz time-domain spectroscopy} 

\begin{figure*}[b]
  \begin{center}
  \includegraphics[width=0.99\textwidth]{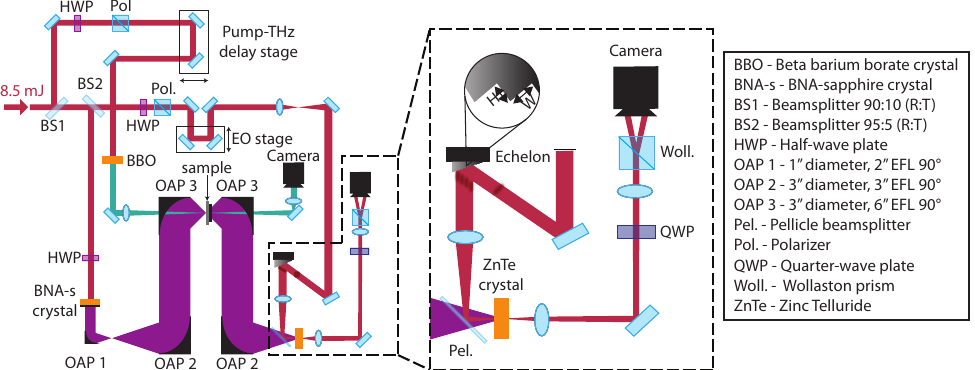}
  \end{center}
\caption{{\bf Schematic of experimental apparatus for THz spectroscopy measurements}. The initial beam is split into two paths, one for pumping the sample and another one for probing the sample and, thus, detecting the initiated dynamic response. Different wavelengths are here represented in different colors with purple representing THz, turquoise representing 400\,nm, and red representing 800\,nm radiation, respectively.}
\label{fig: THz set-up}
\end{figure*}

Figure \ref{fig: THz set-up} illustrates the THz-TDS setup used in our experiments. The design is based on previous experiments \cite{ofori2018toward, ofori2024al, ofori2024dc}. The output of a Ti:sapphire regenerative amplifier (COHERENT Astrella, US) was split: 90\% of the energy used to heat the sample, 9.5\% was used to generate the THz pulse, and the rest was needed for detecting the THz probe. Computer-controlled mechanical shutters were used to block the beams for data acquisition.

The pulses used to heat the films were routed through a half-wave plate (HWP) mounted in a computer-controlled rotation stage and polarizer, to adjust the drive pulse energy, a computer-controlled delay stage, to adjust the arrival time of the drive pulse relative to the THz pulse, and a 0.5 mm thick $\beta$-barium borate (BBO) crystal  (United Crystals, US) to frequency-double the light. This conversion was essential to achieve higher absorbed energies, as copper efficiently absorbs lights with 400\,nm wavelength \cite{creighton1991ultraviolet}. An $f = 75\,$cm focal length lens was used to focus the beam through a hole in an off-axis parabolic reflector (OAP) to the target plane. The spot was subsequently imaged onto a CCD camera (Manta, Allied Vision, Germany). A 2D Gaussian fit to the spot yielded $\sigma_{x}= 373\pm4$ $\mu m$ and $\sigma_{y} = 204\pm4$ $\mu m$ in the sample plane. The highest fluence obtained was  (158.1$\pm$0.5) mJ/cm$^2$. The overlap of the THz and pump beams was confirmed by sending both beams through an empty hole on the sample card and optimizing for the highest transmission.

\subsection*{Generation and detection of the THz probe}

For THz generation, 95\% of the laser energy remaining after splitting the heating pulse was used for optical rectification in a 0.2\,mm thick N-benzyl-2-methyl-4-nitroaniline crystal melted to a sapphire substrate (BNA-s, Terahertz Innovations) \cite{tangen2021comprehensive, zaccardi2021enabling}. The resulting THz beam was then focused using a 1" diameter, 2" effective focal length (EFL) 90$^{\circ}$ OAP, and this focus was relay imaged using 3" diameter, 6" and 3" EFL OAPs onto the target plane. Fig.~1(a) of the main text shows a close-up of the beams that overlapped at the target position. The THz field transmitted through the film was collected and then focused using two OAPs onto a 2\,mm thick (110)-cut zinc telluride crystal (ZnTe).

The remaining energy was used for THz detection via single-shot electro-optic sampling (EOS) \cite{wu1995free, wu1996broadband, nahata1996coherent, Minami2013}. The energy and polarization of the sampling pulse were set using a HWP and polarizer and routed through a separate mechanical stage before being reflected off an echelon mirror (Sodick F.T., Japan). The echelon mirror consists of 120 steps with a width of 150\,$\mu$m and a height of 7.5\,$\mu$m. 

When reflected off the echelon, the incident pulse was split into a series of beamlets, each time-delayed by an amount dictated by the echelon step height. These beamlets were then focused into a zinc tellurde (ZnTe) crystal and overlapped with the THz beam. Through the Pockels effect, the THz field induced a transient birefringence in the ZnTe crystal, whose magnitude is proportional to the THz electric field. This birefringence is encoded as a change in the beamlets' polarization. The beamlets then pass through a quarter-wave plate (QWP) and a Wollaston prism. The Wollaston prism split the transmitted beam into a pair of beams with orthogonal polarizations.
The combination of the QWP and wollaston prism converted the polarization change to an intensity change which was recorded on a charge-coupled device (CCD) camera. A 3-lens system with the QWP and Wollaston prism as inline elements imaged the echelon onto the CCD camera. 

For each condition, we recorded the THz pulse transmitted through the ambient sample, the heated sample, and through the empty hole after the sample was completely ablated. For each target, five ambient and post-mortem measurements were averaged; additionally, we collected a single heated shot per target. We averaged data from three different targets to obtain the values presented in this study.

\subsection*{Sample Characterization}
\subsubsection*{Lateral Characterization}
\begin{figure}[h]
  \centering
  \includegraphics[width=0.5\linewidth]{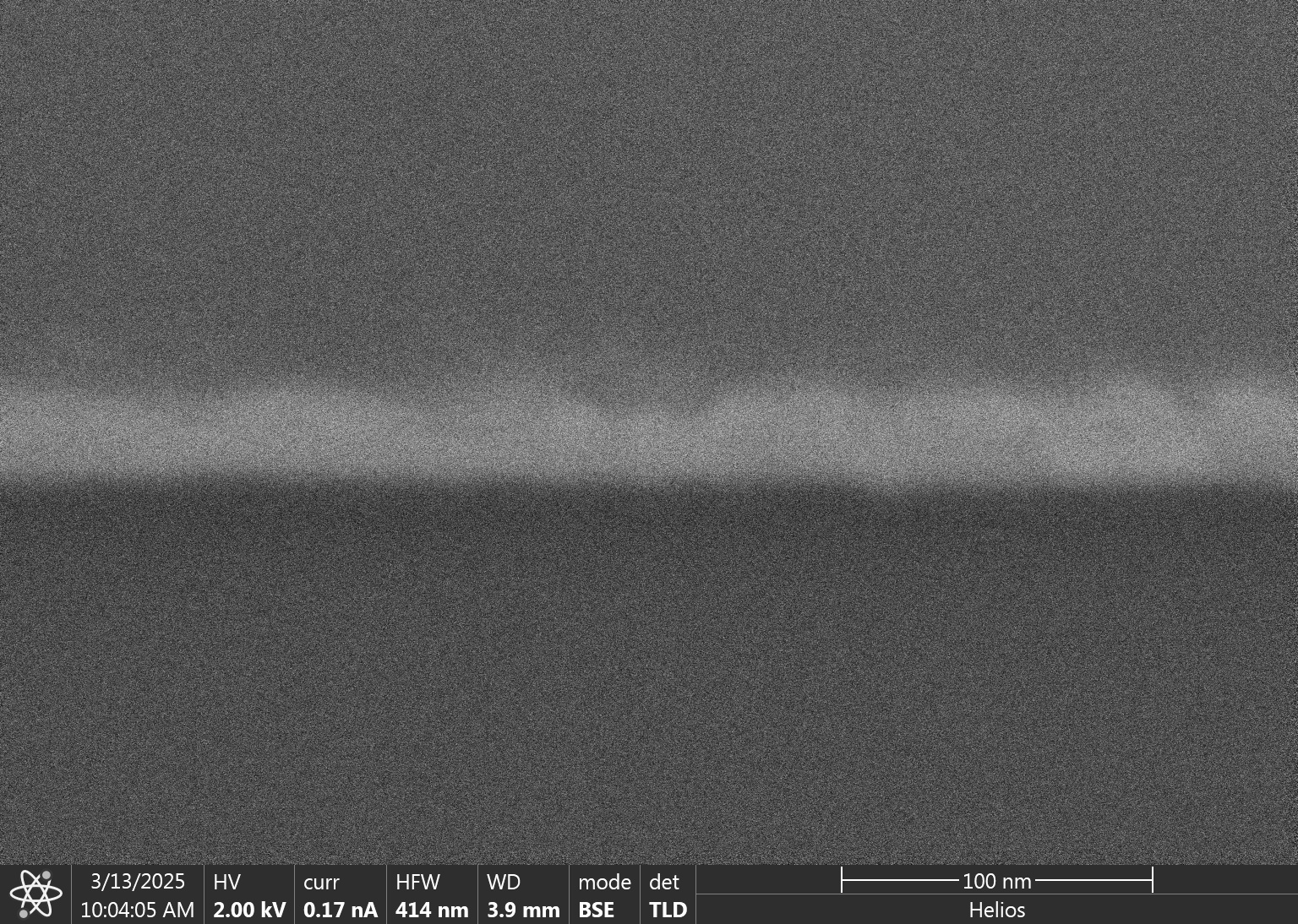}
  \caption{{\bf Analysis of copper grain sizes.} (a) Representative TEM image of a copper sample. Different grains appear as different shades of gray in the image. The scale bar at the bottom right indicates a length of 40\,nm. (b) Distribution of grain sizes in the copper film extracted through a ImageJ analysis. The average grain size was 133\,$\mathrm{nm}^2$.}
  \label{fig:TEM}
\end{figure}

Our copper films were imaged using a FEI Tecnai G2 F20 X-TWIN transmission electron microscopy (TEM), operated at a voltage of 200\,kV. A representative image is shown in Fig.~\ref{fig:TEM}(a). Different grains can be clearly observed through different shades of gray. The image was processed using ImageJ determine the edges of the different grains and thus determine
the different grain areas. A histogram of the different grains is shown in Fig.~\ref{fig:TEM}(b) for the data in Fig.~\ref{fig:TEM}(a). From this we extract an average grain area of 133.8\,$\textrm{nm}^2$. Treating the grains as discs, their effective radius is $\sim$12\,nm,  smaller than the mean free path of conduction electrons in copper, $\sim$40\,nm.

\subsubsection*{Thickness Characterization}
\label{sec: Supp_thickness}

\begin{figure} [h]
 \begin{center}
  \includegraphics[width=\linewidth]{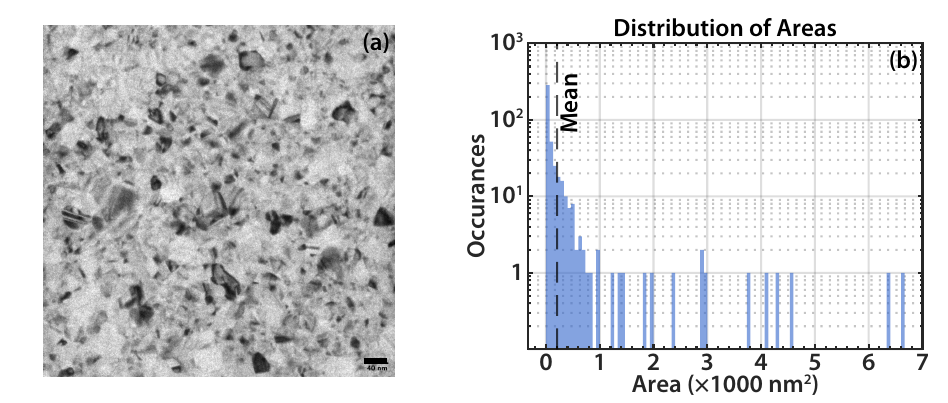} 
 \end{center}
\caption{{\bf FIB/SEM measurement of a 40\,nm copper film.} The figure shows three different layers. The top layer corresponds to a carbon deposition, the middle layer corresponds to the copper film, and the bottom layer corresponds to the underlying silicon wafer.}
\label{fig:SEM}
\end{figure}

We probed different sections of the sample with a Focused Ion Beam Scanning Electron Microscope (FIB/SEM) to determine the thickness of the copper layer in our samples. For this procedure, we used a FEI Helios NanoLab 600i DualBeam system, operating with a current of 0.17\,nA and an acceleration voltage of 2.00\,kV. The thin copper film was coated with a $\sim600$\,nm sacrificial layer of carbon using the Helios FIB. Then a rectangle of 10\,$\mu$m $\times$ 0.5\,$\mu$m was ablated to allow for an observation of the sample's cross-section. This cross-section was captured using the SEM immersion mode at a high voltage of 1\,kV and a current of 43\,pA. A representative cross-section is shown in Fig.~\ref{fig:SEM}. Three distinct layers are observed: the top layer is the carbon deposition; the brighter middle layer is the copper in our sample; the darker bottom layer is the silicon wafer and silicon nitride. Our copper films were produced in two different batches, with SEM measurements giving thicknesses of $30.5\pm3.0$ and $39.3\pm2.7$\,nm for the copper layers.

\subsection*{Two-Temperature Model}

Our measurements were done on a picosecond scale, where each particle species follows equilibrium statistics, but the temperatures of electrons and ions still differ~\cite{Rethfeld2017}.
Ballistic electrons excited by the short laser pulses redistribute heat over their mean free path of approximately $70$\,nm in copper~\cite{hohlfeld_electron_2000}, i.e.\ far beyond the short optical penetration depth, resulting in uniform heating of the film~\cite{maigler2024atomistic, hohlfeld1997time, karna2023direct}.
Furthermore, due to the large laser spot and short timescales of these measurements, we do not expect significant lateral heat transport.

Due to homogeneous conditions, we can model the equilibration of electron and ion temperatures using a two-temperature model~\cite{Anisimov1974}
\begin{align}
  C_{\textrm{e}}(T_{\textrm{e}}) \frac{\partial T_{\textrm{e}}}{\partial t} &= -g_{\textrm{ei}}(T_{\textrm{e}})\left[T_{\textrm{e}}-T_{\textrm{i}}\right]+S(t)\,,
\\[0.5em]
  C_{\textrm{i}} \frac{\partial T_{\textrm{i}}}{\partial t} &= +g_{\textrm{ei}}(T_{\textrm{e}})\left[T_{\textrm{e}}-T_{\textrm{i}}\right]\,,
\end{align}
where $S(t)$ represents the laser heating following a Gaussian temporal profile.
Its strength is determined by the applied fluence.
We use the temperature-dependent electronic heat capacity $C_{\textrm{e}}(T_{\textrm{e}})$ from Ref.~\cite{Lin2008} and a constant ionic heat capacity $C_{\textrm{i}}$ consistent with the Dulong-Petit limit.
The rate of energy transfer between species is governed by the electron-ion coupling parameter $g_{\textrm{ei}}(T_{\textrm{e}})$.
We use the temperature-dependent coupling parameter from Ref.~\cite{migdal2016}, which was previously found to agree well with ultrafast electron diffraction data on copper~\cite{MoCu}.

\subsection*{TTM Molecular Dynamics}

Coupled TTM-MD modeling \cite{Schaefer2002,Ivanov2003} was performed using the LAMMPS software package~\cite{LAMMPS} with the implementation proposed in Refs.~\cite{duffy2007,pisarev2014}.
Moreover, an adaptive electron voxel checking algorithm to account for empty and non-empty cells of the continuous background grid as well as a temperature-dependent coupling parameter were implemented~\cite{maigler2024atomistic}.

The governing equation tracking the trajectory of the $i$-th atom is expressed as
\begin{equation}
  m_i \cdot \frac{\partial \vec{u_i}}{\partial t} = \vec{F_i}(t) - \gamma \vec{u_i} + \hat{\vec{F}}(t) \,,
  \label{eq:ttm_md1}
\end{equation}
where $m_i$, $\vec{u_i}$, $\vec{F_i}$, $\gamma_i$, and $\hat{\vec{F}}$ denote mass, velocity, exerted force, friction coefficient, and stochastic exerted force, respectively.
The subscript $i$ indicates the quantities per-atom.
The interatomic forces $\vec{F_i}$ were derived from the embedded-atom potential of Ref.~\cite{sheng2011} chosen for its reliable thermal and mechanical properties.
The friction coefficient
\begin{equation}
  \gamma = \frac{g_{\textrm{ei}} \cdot V \cdot M_V}{3N k_B}\,,
  \label{eq:ttm_md2}
\end{equation}
accounts for the interaction between the lattice/ion and the electronic subsystems.
Here, $V$ is the voxel volume, $M_V=\sum_i^V m_i$ is the total mass of atoms in this voxel, $N$ indicates the number of atoms, and $k_\textrm{B}$ is the Boltzmann constant.

\newpage
\bibliographystyle{ScienceAdvances}
\bibliography{apssamp}   

\vspace{11mm}
\noindent {\bf Acknowledgments:}
The authors thank G.~Glenn, A.~Marret, and J.~Gama~Vazquez for useful discussions on the theoretical modeling. B.R.\ thanks the HEDS Division at SLAC for its hospitality during multiple visits.
{\bf Funding:}
This work was in part supported by the DOE Office of Science, Fusion Energy Science under FWPs 100182 and 100866, and by the Department of Energy Laboratory Directed Research and Development program at SLAC National Accelerator Laboratory under contract no.\ DE-A02-76SF00515 and as part of the Panofsky Fellowship awarded to B.O.O.
T.H.\ and B.R.\ gratefully acknowledge the support from the Deutsche Forschungsgemeinschaft (DFG, German Research Foundation) through the SFB/TRR-173-268565370 ‘Spin+X’ (Project No.\ A08 and INF) and the Allianz für Hochleistungsrechnen Rheinland-Pfalz for providing computing resources through project STREMON on the Elwetritsch high-performance computing cluster.
{\bf Author contributions:} 
E.R.T., S.H.G., and B.O.O.\ conceived the study. E.R.T. and B.O.O.\ designed the samples. E.R.T., M.I., E.R.S., and B.O.O.\ conducted the experiments. T.H.\ and M.M.\ performed the simulations. E.R.T.\ analyzed the data. T.H.\ 
developed the theoretical framework used for interpretation. E.R.T., T.H., A.B., M.M., D.O.G., M.Z.M, B.R., S.H.G., and B.O.O.\ interpreted the results. E.R.T, T.H., D.O.G., B.R., and B.O.O.\ wrote the manuscript with input from all authors. 
{\bf Competing interests:}
The authors declare that they have no competing interests. 
{\bf Data and material availability:}
The datasets generated and analyzed
during this study are available from the corresponding authors upon reasonable request.
\clearpage

\clearpage
\clearpage
\clearpage
\clearpage
\clearpage
\clearpage
\end{document}